\documentclass[twocolumn,amssymb,aps,showpacs,longbibliography]{revtex4-1}
\usepackage{graphicx}
\usepackage[colorlinks]{hyperref}
\usepackage{amsbsy}

\begin{document}

\title{Collective excitations on a surface of topological insulator}

\author{D.\,K. Efimkin${}^1$}
\author{Yu.\,E. Lozovik${}^{1,2}$}\email{lozovik@isan.troitsk.ru}
\author{A.\,A. Sokolik${}^1$}

\affiliation{${}^1$Institute for Spectroscopy RAS, 142190, Troitsk, Moscow Region, Russia\\
${}^2$Moscow Institute of Physics and Technology, 141700, Moscow, Russia}%

\begin{abstract}
We study collective excitations in a helical electron liquid on a surface of three-dimensional topological insulator.
Electron in helical liquid obeys Dirac-like equation for massless particless and direction of its spin is strictly
determined by its momentum. Due to this spin-momentum locking, collective excitations in the system manifest themselves
as coupled charge- and spin-density waves. We develop quantum field-theoretical description of spin-plasmons in helical
liquid and study their properties and internal structure. Value of spin polarization arising in the system with excited
spin-plasmons is calculated. We also consider the scattering of spin-plasmons on magnetic and nonmagnetic impurities
and external potentials, and show that the scattering occurs mainly into two side lobes. Analogies with Dirac electron
gas in graphene are discussed.
\end{abstract}


\maketitle

\section{Introduction}
Topological insulator is a new class of solids with nontrivial topology, intrinsic to its band structure. Theoretical
and experimental studies of topological insulators grow very rapidly in recent years (see \cite{Hasan,QiZhang} and
references therein). Three-dimensional topological insulators are insulating in the bulk, but have gapless surface
states with numerous unusual properties. These states are topologically protected against time-reversal invariant
disorder. When gap is opened in surface states by time-reversal or gauge symmetry breaking, a spectacular
magnetoelectric effect arises \cite{QiHughesZhang,EssinMooreVanderbilt}.

Recently a ``new generation'' of three-dimensional topological insulators (the binary compounds $\mathrm{Bi}_2
\mathrm{Se}_3$, $\mathrm{Bi}_2 \mathrm{Te}_3$ and other materials), retaining topologically protected behavior at room
temperatures, were predicted and studied experimentally \cite{Chen,Hsieh,Xia}. Band structure of the surface states of
these materials contain a single Dirac cone, where electrons obey two-dimensional Dirac equation for massless
particles. Direction of electron spin in these states is strictly determined by their momentum, so these states can be
called as ``helical'' ones. Surface of topological insulator can be chemically doped, forming charged helical liquid.
The spin-momentum locking leads to interesting transport phenomena including coupled diffusion of spin and charge
\cite{Burkov}, inverse galvano-magnetic effect (generation of spin polarization by electric current) \cite{Cucler} and
giant spin rotation on an interface between normal metal and topological insulator \cite{YokoyamaTanakaNagaosa}. The
spin-momentum locking offers numerical opportunities for various spintronic applications.

Collective excitations (plasmons) in helical liquid on the surface of topological insulator was considered in
\cite{Raghu}. It was shown that due to spin-momentum locking responses of charge and spin densities to external
electromagnetic field are coupled to each other. Therefore the plasmons in the system should manifest themselves as
coupled charge- and spin-density waves and can be called ``spin-plasmons''. In \cite{Appelbaum} application of
spin-plasmons in spin accumulator device was proposed. Also surface plasmon-polaritons under conditions of topological
magnetoelectric effect were considered in \cite{Karch}.

Properties of the states on a surface of three-dimensional topological insulator are similar to those of electrons in
graphene. Graphene is unique two-dimensional carbon material with extraordinary electronic properties
\cite{Graphene1,Graphene2,CastroNeto}. Its band structure contains two Dirac cones with electrons behaving as massless
Dirac particles in their vicinities. Graphene is a perspective material for nanoelectronics due to large carrier
mobilities at room temperature. Electronic interactions and collective excitations in graphene have been extensively
studied (see \cite{KotovUchoaPereiraCastroNetoGuinea} and references therein). In particular, the properties of
plasmons \cite{HwangDasSarma,WunchStauberSolsGuinea} and hybrid plasmon-photon \cite{BludovVasilevskiyPeres} and
plasmon-phonon \cite{HwangSensarmaDasSarma,LiuWillis} modes were investigated theoretically and experimentally.  It was
realized recently that graphene is a fertile ground for quantum plasmonics \cite{KoppensChangGarciaAbajo} due to very
small plasmon damping.

In this article, we develop quantum field-theoretical formalism to describe plasmons in graphene and spin-plasmons of
on a surface of three-dimensional topological insulator based on random phase approximation. Problems of excitation,
manipulation, scattering and detection of single plasmons can be conveniently considered using this approach. Thus,
this approach can be especially useful for problems of plasmon quantum optics and quantum plasmonics. We use our
approach to study internal structure and properties of spin-plasmons in a helical liquid.

The rest of this paper is organized as follows. In section 2, we present a brief description of electronic states on a
surface of topological insulator and in graphene. In section 3, quantum field-theoretical description of plasmons in
Dirac electron gas is given. In section 4, we consider internal structure of plasmons. In section 5, different
consequences of spin-momentum locking are considered. Scattering of plasmons on impurities and external potentials is
considered in section 6, and section 7 is devoted to conclusions.

\section{Dirac electrons}
The low-energy effective Hamiltonians for electrons in helical liquid \cite{Zhang} and graphene \cite{CastroNeto} are
\begin{eqnarray}
\mbox{helical liquid:}&&H_{0}=v_\mathrm{F}(p_x\sigma_y-p_y\sigma_x),\label{HamHelicalLiquid}\\
\mbox{graphene:}&&H_{0}=v_\mathrm{F}(p_x\sigma_x+p_y\sigma_y),\label{HamGraphene}
\end{eqnarray}
where the Fermi velocities of electrons $v_\mathrm{F}$ are $6.2\times10^5\:\mbox{m/s}$ for the topological insulator
$\mathrm{Bi}_2\mathrm{Se}_3$ and $10^6\:\mbox{m/s}$ for graphene; the Pauli matrices $\sigma_x$ and $\sigma_y$ act in
the spaces of electron spin projections (helical liquid) or sublattices (graphene). The eigenfunctions of
(\ref{HamHelicalLiquid})--(\ref{HamGraphene}) can be written as
$e^{i\mathbf{p}\cdot\mathbf{r}}|f_{\mathbf{p}\gamma}\rangle/\sqrt{S}$, where $S$ is the system area and
$|f_{\mathbf{p}\gamma}\rangle$ is the spinor part of the eigenfunction, corresponding to electron with momentum
$\mathbf{p}$  from conduction ($\gamma=+1$) or valence ($\gamma=-1$) band:
\begin{eqnarray}
\mbox{helical liquid:}&&|f_{\mathbf{p}\gamma}\rangle=\frac1{\sqrt2}\left(\begin{array}{c}
e^{-i\varphi_\mathbf{p}/2}\\
i\gamma e^{i\varphi_\mathbf{p}/2}\end{array}\right),\\
\mbox{graphene:}&&|f_{\mathbf{p}\gamma}\rangle=\frac1{\sqrt2}\left(\begin{array}{c}e^{-i\varphi_{\mathbf{p}}/2}\\
\gamma e^{i\varphi_{\mathbf{p}}/2}\end{array}\right)
\end{eqnarray}
(here and below we assume $\hbar=1$). Another difference between helical liquid on a surface of topological insulator
and electron liquid in graphene is additional fourfold degeneracy $g=4$ of electronic states in graphene by two spin
projections and two nonequivalent valleys.

The value of electron spin in helical liquid and of pseudospin graphene in the state $|f_{\mathbf{p}\gamma}\rangle$ is
\begin{eqnarray}
\mbox{helical liquid:}&&\langle f_{\mathbf{p}\gamma}|\boldsymbol\sigma|f_{\mathbf{p}\gamma}\rangle=\gamma
[\hat\mathbf{z}\times\hat\mathbf{p}],\\
\mbox{graphene:}&&\langle f_{\mathbf{p}\gamma}|\boldsymbol\sigma|f_{\mathbf{p}\gamma}\rangle= \gamma \hat\mathbf{p},
\end{eqnarray}
where $\hat\mathbf{p}$ and $\hat\mathbf{z}$ are unit vectors directed along the momentum $\mathbf{p}$ and the $z$-axis
respectively. We see that, in helical liquid, the spin of electron lies in the system plane and makes an angle
$90^\circ$ (in counterclockwise direction in the conduction band and inversely in the valence band) with its momentum.
In graphene, the sublattice pseudospin of electron is directed along its momentum in conduction band and opposite to it
in the valence band. Physically, a definite direction of the pseudospin in the system plane corresponds to definite
phase shift between electron wave functions on different sublattices.

A starting point for quantum field-theoretical consideration of plasmons on the surface of topological insulator and in
graphene is the many-body Hamiltonian of electrons with Coulomb interaction between them:
\begin{eqnarray}
H=\sum_{\mathbf{p}\gamma}\xi_{p\gamma}a_{\mathbf{p}\gamma}^+a_{\mathbf{p}\gamma}+
\frac{1}{2S}\sum_{\mathbf{q}\mathbf{p}\mathbf{p}^\prime}\sum_{\gamma_1\gamma_2}\sum_{\gamma_1^\prime\gamma_2^\prime}
V_q\nonumber\\ \times\langle f_{\mathbf{p}+\mathbf{q},\gamma_1^\prime}|f_{\mathbf{p}\gamma_1}\rangle \langle
f_{\mathbf{p}^\prime-\mathbf{q},\gamma_2^\prime}|f_{\mathbf{p}^\prime\gamma_2}\rangle\nonumber\\
\times a_{\mathbf{p+q},\gamma_1^\prime}^+a^+_{\mathbf{p}^\prime-\mathbf{q},\gamma_2^\prime}
a_{\mathbf{p}^\prime\gamma_2}a_{\mathbf{p}\gamma_1},\label{ManyBodyHam}
\end{eqnarray}
where $a_{\mathbf{p}\gamma}$ is the destruction operator for electron with momentum $\mathbf{p}$ from the band
$\gamma$, $\xi_{p\gamma}=\gamma v_\mathrm{F}p-\mu$ is its kinetic energy measured from the chemical potential $\mu$ and
$V_q=2\pi e^2/\varepsilon q$ is the two-dimensional fourier transform of Coulomb interaction potential screened by
surrounding three-dimensional medium with a dielectric susceptibility $\varepsilon$.

\section{Description of plasmons}
To investigate the properties of plasmons in Dirac electron gas, we use equation-of-motion approach \cite{Sawada}. We
treat a plasmon as a composite quasiparticle, consisting of electron-hole pairs with the common total momentum
$\mathbf{q}$. Thus, the creation operator for plasmon with momentum $\vec{q}$ can be written as:
\begin{equation}
Q_{\vec{q}}^+=\sum_{\vec{p}\gamma\gamma^\prime}C_{\vec{pq}}^{\gamma^\prime\gamma}
a_{\vec{p}+\vec{q},\gamma^\prime}^+a_{\mathbf{p}\gamma}.\label{OperatorPlasmon}
\end{equation}
Here the coefficients $C_{\vec{pq}}^{\gamma^\prime\gamma}$ are the weights of intraband ($\gamma=\gamma^\prime$) and
interband ($\gamma\neq\gamma^\prime$) single-particle transitions, contributing to the wave function of plasmon.

The plasmon creation operator should obey Heisenberg equation of motion
\begin{equation}
[H,Q_{\vec{q}}^+]=\Omega_q Q_{\vec{q}}^+,\label{eqPlasm}
\end{equation}
where $\Omega_q$ is plasmon frequency. We start from equation of motion for single electron-hole pair, which can be
derived using (\ref{ManyBodyHam}):
\begin{eqnarray}
\left[H,a_{\mathbf{p}+\mathbf{q},\gamma^\prime}^+a_{\mathbf{p}\gamma}
\right]\nonumber\\=(\xi_{\mathbf{p}+\mathbf{q},\gamma^\prime}-\xi_{\mathbf{p}\gamma})
a_{\mathbf{p}+\mathbf{q},\gamma^\prime}^+a_{\mathbf{p}\gamma}+
\frac1S\sum_{\mathbf{q}^\prime}V_{q^\prime}\rho^+_{\mathbf{q}^\prime}\nonumber\\
\times\sum_{\gamma^{\prime\prime}}\left(\langle
f_{\mathbf{p}+\mathbf{q}-\mathbf{q}^\prime,\gamma^{\prime\prime}}|f_{\mathbf{p}+\mathbf{q},\gamma^{\prime}} \rangle
a_{\mathbf{p}+\mathbf{q}-\mathbf{q}^\prime,\gamma^{\prime\prime}}^+a_{\mathbf{p}\gamma}\right.\nonumber\\
\left.-\langle f_{\mathbf{p}\gamma}|f_{\mathbf{p}+\mathbf{q}^\prime,\gamma^{\prime\prime}}\rangle
a_{\mathbf{p}+\mathbf{q},\gamma^{\prime}}^+a_{\mathbf{p}+\mathbf{q}^\prime,\gamma^{\prime\prime}}\right),
\label{eqSingle}
\end{eqnarray}
where we have introduced the Fourier transform of the density operator:
\begin{equation}
\rho^+_{\mathbf{q}}=\sum_{\mathbf{p}\gamma\gamma'}\langle f_{\mathbf{p}+\mathbf{q},\gamma'}|f_{\mathbf{p}\gamma}\rangle
a^+_{\mathbf{p}+\mathbf{q},\gamma'}a_{\mathbf{p}\gamma}.\label{rhoDensity}
\end{equation}

The right-hand side of the equation (\ref{eqSingle}) contains products of four fermionic operators. To reduce it to
products of two fermionic operators we use the random phase approximation (RPA); its applicability will be discussed
below. According to RPA \cite{Sawada}, the operator products in the last two lines of (\ref{eqSingle}) can be replaced
by their average values in ground state $|0\rangle$ of the system:
\begin{equation}
a^+_{\mathbf{p}^\prime\gamma^\prime}a_{\mathbf{p}\gamma}\;\rightarrow\;\langle0|
a^+_{\mathbf{p}^\prime\gamma^\prime}a_{\mathbf{p}\gamma}|0\rangle=
\delta_{\mathbf{p}\mathbf{p}^\prime}\delta_{\gamma\gamma^\prime} n_{\mathbf{p}\gamma},\label{RPA}
\end{equation}
where $n_{\mathbf{p}\gamma}$ is the occupation number for electronic states with momentum $\mathbf{p}$ from the band
$\gamma$. For electron-doped Dirac liquid at $T=0$ (see also the remark at the end of this section), we have
$n_{\mathbf{p}+}=\Theta(p_\mathrm{F}-|\mathbf{p}|)$, $n_{\mathbf{p}-}=1$, where $p_\mathrm{F}=\mu/v_\mathrm{F}$ is the
Fermi momentum. In the case of hole doping, all characteristics of plasmons are the same due to electron-hole symmetry.

Thus, the equation of motion (\ref{eqSingle}) for electron-hole pair with taking into account the RPA assumption
(\ref{RPA}) takes the form:
\begin{eqnarray}
\left[H,a_{\mathbf{p}+\mathbf{q},\gamma^\prime}^+a_{\mathbf{p}\gamma}
\right]=(\xi_{\mathbf{p}+\mathbf{q},\gamma^\prime}-\xi_{\mathbf{p}\gamma})
a_{\mathbf{p}+\mathbf{q},\gamma^\prime}^+a_{\mathbf{p}\gamma}\nonumber\\
+\frac{V_q}{S} \rho^+_{\mathbf{q}} \langle f_{\mathbf{p}\gamma}|f_{\mathbf{p}+\mathbf{q},\gamma^{\prime}} \rangle
\left( n_{\mathbf{p}\gamma}-n_{\mathbf{p}+\mathbf{q},\gamma'}\right).\label{eqehRPA}
\end{eqnarray}
Combining the definition of plasmon creation operator (\ref{OperatorPlasmon}) with the equation of motion for plasmon
(\ref{eqPlasm}) and single electron-hole pair (\ref{eqehRPA}), we obtain the system of equations for the coefficients
$C_{\vec{pq}}^{\gamma^\prime\gamma}$:
\begin{eqnarray}
(\Omega_q+\xi_{\mathbf{p}\gamma}-\xi_{\mathbf{p}+\mathbf{q},\gamma^\prime})
C^{\gamma^\prime\gamma}_{\mathbf{pq}}=\frac{V_q}{S} \langle
f_{\mathbf{p}+\mathbf{q},\gamma^{\prime}}|f_{\mathbf{p}\gamma}\rangle\nonumber\\
\times\sum_{\mathbf{p}^\prime\tau\tau'} C_{\mathbf{p}^\prime \mathbf{q}}^{\tau'\tau}\langle
f_{\mathbf{p}^\prime\tau}|f_{\mathbf{p}^\prime+\mathbf{q},\tau^{\prime}} \rangle \left(
n_{\mathbf{p}^\prime\tau}-n_{\mathbf{p}^\prime+\mathbf{q},\tau^\prime}\right).\label{eqPlasm1}
\end{eqnarray}

Introducing infinitesimal damping into the plasmon frequency and denoting
\begin{equation}
N_{\mathbf{q}}=\frac{V_q}{S}\sum_{\mathbf{p}\tau\tau'}C_{\mathbf{p}\mathbf{q}}^{\tau'\tau}\langle
f_{\mathbf{p}\tau}|f_{\mathbf{p}+\mathbf{q},\tau^{\prime}}
\rangle\left(n_{\mathbf{p}\tau}-n_{\mathbf{p}+\mathbf{q},\tau^\prime}\right),\label{N}
\end{equation}
we can find the plasmon wave function from (\ref{eqPlasm1}) as:
\begin{eqnarray}
C_{\mathbf{p}\mathbf{q}}^{\gamma'\gamma}=\frac{\langle f_{\mathbf{p}+\mathbf{q},\gamma'}|f_{\mathbf{p}\gamma}\rangle
N_{\mathbf{q}}}{\Omega_q+\xi_{\mathbf{p}\gamma}-\xi_{\mathbf{p}+\mathbf{q},\gamma'}+i\delta}.\label{C}
\end{eqnarray}

The normalization factor $N_\mathbf{q}$ can be determined from the commutation relation for plasmon operators, which
should be satisfied on the average in the ground state:
\begin{equation}
\langle0|[Q_{\mathbf{q}},Q_{\mathbf{q}'}^+]|0\rangle=\delta_{\mathbf{q}\mathbf{q}'}.\label{norm1}
\end{equation}
Substituting (\ref{OperatorPlasmon}) in (\ref{norm1}), we get
\begin{equation}
\sum_{\gamma\gamma'}D_{\gamma'\gamma}= 1,\label{norm}
\end{equation}
where the total weights of intraband ($D_{++}$) and interband ($D_{+-}+D_{-+}=1-D_{++}$) electron transitions,
contributing to the plasmon wave function (\ref{C}), are:
\begin{equation}
D_{\gamma'\gamma}=\sum_{\mathbf{p}}\left|C_{\mathbf{p}\mathbf{q}}^{\gamma'\gamma}\right|^2
(n_{\mathbf{p}\gamma}-n_{\mathbf{p}+\mathbf{q},\gamma'}).\label{D}
\end{equation}

To find the plasmon frequency $\Omega_q$, we can substitute (\ref{C}) in (\ref{N}), arriving to the standard RPA
equation (see it for the helical liquid \cite{Raghu} and for graphene \cite{HwangDasSarma,WunchStauberSolsGuinea}):
\begin{equation}
1-V_q\Pi(q,\Omega_q)=0,\label{eqDispersion}
\end{equation}
where the polarization operator of Dirac electron gas is introduced:
\begin{eqnarray}
\Pi(q,\omega)=\frac{g}S\sum_{\mathbf{p}\gamma\gamma'}\left|\langle
f_{\mathbf{p}+\mathbf{q},\gamma'}|f_{\mathbf{p}\gamma}\rangle\right|^2\nonumber\\
\times\frac{n_{\mathbf{p}\gamma}-n_{\mathbf{p}+\mathbf{q},\gamma'}}
{\omega+\xi_{\mathbf{p}\gamma}-\xi_{\mathbf{p}+\mathbf{q},\gamma'}+i\delta}.\label{Pol}
\end{eqnarray}
The degeneracy factor $g$ is 1 for helical liquid on the surface of topological insulator and 4 for graphene. The
angular factors $\langle f_{\mathbf{p}+\mathbf{q},\gamma'}|f_{\mathbf{p}\gamma}\rangle$ are specific to chiral Dirac
electron and arise in (\ref{Pol}) as a result of summation over spinor components of electron wave function.

The random phase approximation becomes exact in the limit of small values of dimensional parameter $r_\mathrm{s}$,
defined as a ratio of characteristic Coulomb interaction energy to kinetic energy. For the gas of massless particles,
$r_\mathrm{s}$ does not depend on electron density and equals to $r_\mathrm{s}=e^2/\varepsilon v_\mathrm{F}$
(effectively, $\varepsilon=(\varepsilon_1+\varepsilon_2)/2$ when two-dimensional electron layer is surrounded by two
dielectric half-spaces with susceptibilities $\varepsilon_1$ and $\varepsilon_2$). For $\mathrm{Bi}_2\mathrm{Se}_3$,
$r_\mathrm{s}=0.09$ with $\epsilon=40$ for dielectric half-space, and applicability of the RPA is well established (the
value of $r_\mathrm{s}$ for another material $\mathrm{Bi}_2\mathrm{Te}_3$ is close to that for
$\mathrm{Bi}_2\mathrm{Se}_3$). In the case of graphene, $r_\mathrm{s}$ is not small, but applicability of the RPA can
be established due to smallness of the parameter $1/g$, leading to selection of bubble diagrams (see \cite{CastroNeto}
and references therein, and also the work \cite{Apenko}).

Maximal achievable amounts of doping of helical liquid on a surface of $\mathrm{Bi}_2\mathrm{Se}_3$ are
$\mu\sim0.3\:\mbox{eV}$ \cite{Zhang}, therefore at room temperature it can be degenerate electron liquid. Hence we
assume that $T=0$ in all calculations below.

\section{Wave function of plasmon}
\begin{figure}
\includegraphics[width=0.9\columnwidth]{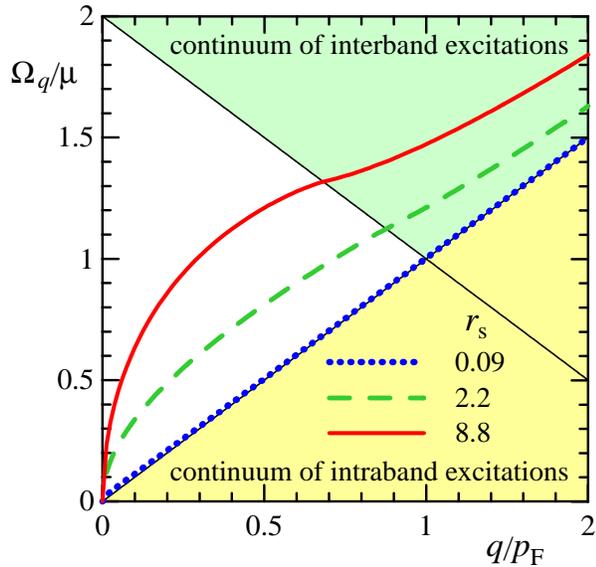}
\caption{Plasmon dispersions $\Omega_q$ at different values of $r_\mathrm{s}$. Continuums of intraband
($\omega<v_\mathrm{F}q$) and interband ($\omega>2\mu-v_\mathrm{F}q$) single-particle excitations are shaded;
$p_\mathrm{F}=\mu/v_\mathrm{F}$ is the Fermi momentum, $\mu$ is the chemical potential.}\label{Fig1}
\end{figure}

Plasmon dispersions $\Omega_q$, calculated numerically from (\ref{eqDispersion})--(\ref{Pol}) at various
$r_\mathrm{s}$, are plotted in Fig.~\ref{Fig1}. We also present results for suspended graphene with $r_\mathrm{s}=8.8$
($\varepsilon=1$) and for graphene embedded into $\hbox{SiO}_2$ environment with $r_\mathrm{s}=2.2$ ($\varepsilon=4$).
For convenience, the degeneracy factor $g=4$ is included here into $r_\mathrm{s}$, since the plasmon dispersion in RPA
(\ref{eqDispersion}) depend only on the combination $gr_\mathrm{s}$.

It is seen that for small values of $r_\mathrm{s}$ (the case of topological insulator) the plasmon dispersion law
approaches very closely to the upper border of the continuum of intraband single-particle transitions
($\omega<v_\mathrm{F}q$). On the contrary, at moderate and large $r_\mathrm{s}$, plasmon has well-defined square-root
dispersion in the long-wavelength range. At $q\approx p_\mathrm{F}$, the plasmon enters the continuum of interband
single-particle transitions ($\omega>2\mu-v_\mathrm{F}q$). Inside the single-particle continuum, energy and momentum
conservation laws allow energy transfer between plasmon and single-particle excitations, so the plasmon acquires a
finite lifetime.

\begin{figure}
\includegraphics[width=0.9\columnwidth]{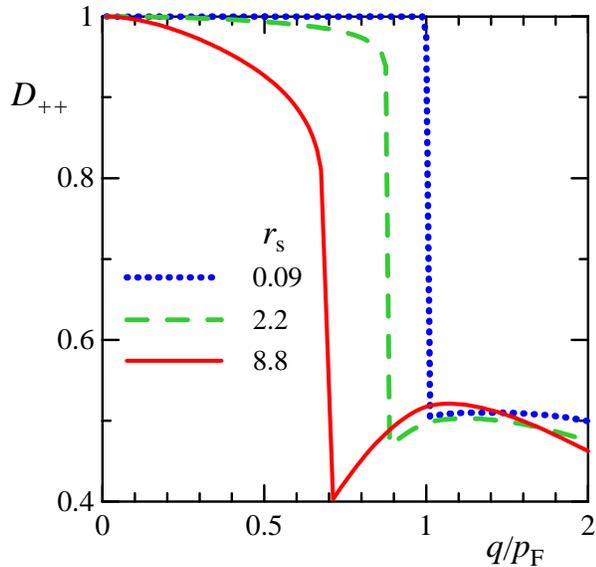}
\caption{Weight $D_{++}$ (\ref{D}) of intraband electron-hole transitions in plasmon wave function at different values
of $r_\mathrm{s}$.}\label{Fig2}
\end{figure}

The internal structure of plasmons in Dirac electron gas can be investigated by looking at total weights (\ref{D}) of
intra- and interband transitions in its wave function. The weight $D_{++}$ of intraband single-particle transitions is
plotted in Fig.~\ref{Fig2}. It is seen that the undamped plasmon for all values of parameter $r_\mathrm{s}$ consists
mainly of intraband transitions, thus an influence of valence band on its properties is rather weak. When the plasmon
enters the single-particle continuum, inter- and intraband transitions start to contribute almost equally to its wave
function, but the plasmon undergoes strong Landau damping.

\begin{figure}
\includegraphics[width=\columnwidth]{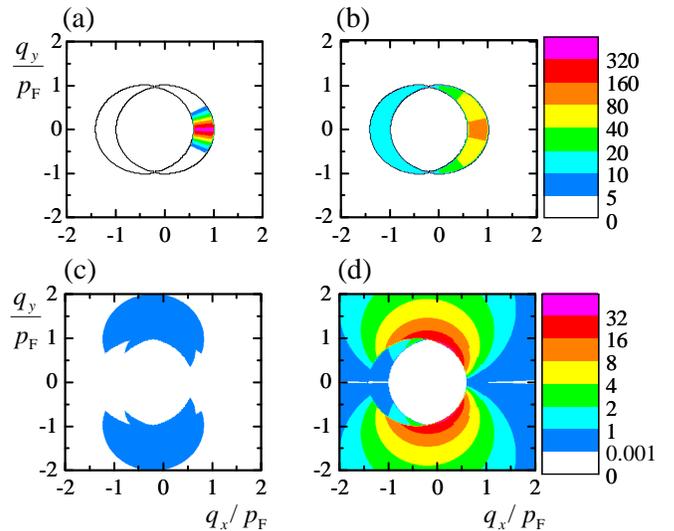}
\caption{Squared modulus of plasmon wave function $|C^{\gamma'\gamma}_{\mathbf{p}\mathbf{q}}|^2$ on the plane of the
hole momentum $\mathbf{p}$ at $q=0.4p_\mathrm{F}$. Top row: intraband channel $|C^{++}_{\mathbf{p}\mathbf{q}}|$ at
$r_\mathrm{s}=0.09$ (a) and $r_\mathrm{s}=8.8$ (b). Bottom row: interband channel
$|C^{+-}_{\mathbf{p}\mathbf{q}}|+|C^{-+}_{\mathbf{p}\mathbf{q}}|$ at $r_\mathrm{s}=0.09$ (c) and $r_\mathrm{s}=8.8$
(d).}\label{Fig3}
\end{figure}

The detailed picture of the plasmon wave function in momentum space $C_{\mathbf{pq}}^{\gamma'\gamma}$, showing the
distribution of contributions of intraband $|C_{\vec{pq}}^{++}|^2$ or interband
$|C_{\vec{pq}}^{+-}|^2+|C_{\vec{pq}}^{-+}|^2$ electron-hole pairs into wave function of plasmon with given momentum $q$
is presented in Fig.~\ref{Fig3}. The results are calculated for $q=0.4p_\mathrm{F}$ at $r_\mathrm{s}=0.09$ and $8.8$.
For other values of plasmon momentum $q$, the results are qualitatively similar. It is seen that for small values of
$r_\mathrm{s}$ (the case of topological insulator) the distribution of intraband contributions is very sharply peaked
in the forward direction, whereas the contribution of interband transitions is negligible. It can be understood from
the reason that the plasmon dispersion is very close to the single-particle continuum (see Fig.~\ref{Fig1}) and thus
the plasmon itself behave almost as single intraband electron-hole transition.

At large values of parameter $r_\mathrm{s}$ (the case of suspended graphene), the broad range of electron intraband
transitions in momentum space contribute to plasmon. The weight of interband transitions is small but not negligible.
Contributions of interband transitions form two side lobes, since the angular factor $\langle
f_{\mathbf{p}+\mathbf{q},\pm}|f_{\mathbf{p}\mp}\rangle$ suppresses interband forward scattering.

\section{Charge- and spin-density waves}
When a spin-plasmon is excited in the helical liquid, anisotropic distribution of electron-hole pairs of the type,
depicted in Fig.~\ref{Fig3}, arises. This distribution is shifted towards the plasmon momentum $\mathbf{q}$. Due to the
spin-momentum locking, the system should acquire a total nonzero spin polarization, perpendicular to $\mathbf{q}$. A
similar situation occurs in the current-carrying state of the helical liquid, which turns out to be spin-polarized
\cite{Cucler}.

\begin{figure}
\includegraphics[width=0.9\columnwidth]{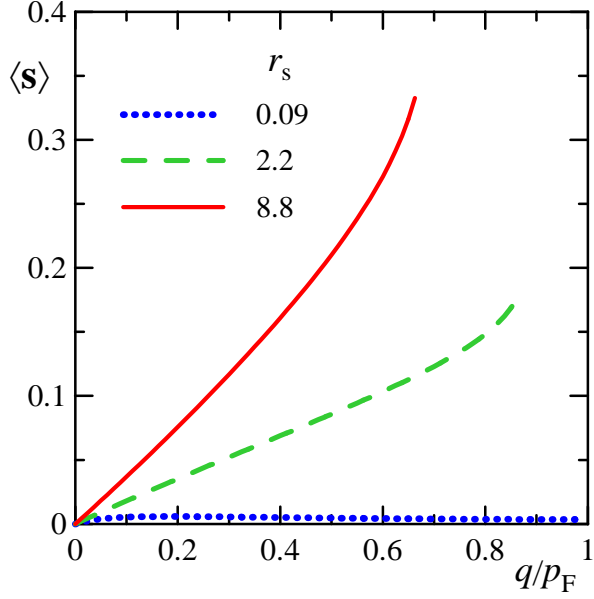}
\caption{In-plane transverse spin polarization of the system $\langle\mathbf{s}\rangle$ (\ref{st}) in the state with
single spin-plasmon excited as function of its momentum $q$ at different values of $r_\mathrm{s}$.}\label{Fig4}
\end{figure}

Average of spin polarization in the state $|1_\mathbf{q}\rangle=Q^+_\mathbf{q}|0\rangle$ with a single plasmon excited
is $\langle\mathbf{s}\rangle=\langle1_{\mathbf{q}}|\boldsymbol\sigma/2|1_{\mathbf{q}}\rangle$ and can be calculated
using (\ref{OperatorPlasmon}) as:
\begin{eqnarray}
\langle\mathbf{s}\rangle=\frac12\sum_{\mathbf{p}\gamma\gamma'\tau}\left[\langle
f_{\mathbf{p}+\mathbf{q},\gamma'}|\boldsymbol\sigma|f_{\mathbf{p}+\mathbf{q},\tau}\rangle
C^{\tau\gamma}_{\mathbf{p}\mathbf{q}}\vphantom{C^{\gamma'\tau}_{\mathbf{p}\mathbf{q}}}\right.\nonumber\\
\left.-C^{\gamma'\tau}_{\mathbf{p}\mathbf{q}}\langle
f_{\mathbf{p}\tau}|\boldsymbol\sigma|f_{\mathbf{p}\gamma}\rangle\right]
\left(C_{\mathbf{p}\mathbf{q}}^{\gamma'\gamma}\right)^*
(n_{\mathbf{p}\gamma}-n_{\mathbf{p}+\mathbf{q},\gamma'}).\label{st}
\end{eqnarray}
If $\mathbf{q}$ is parallel to $\mathbf{e}_x$, only the $y$-component of $\langle\mathbf{s}\rangle$ is nonzero. Its
dependence on $q$ for undamped plasmons at various $r_\mathrm{s}$ is plotted in Fig.~\ref{Fig4}. At large enough
$r_\mathrm{s}$, the spin polarization of the system is comparable to the whole spin of a single electron.

Note that in the case of graphene the isospin polarization of the system appears instead of spin polarization. Isospin
polarization corresponds to nonzero average phase shift between wave function of electrons on different sublattices. On
the contrary, in pseudospin-unpolarized state this shift is zero on the average.

In ordinary electronic system without spin-momentum coupling caused by spin-orbit interaction, plasmon manifests itself
as charge-density wave. In helical liquid on surface of topological insulator, due to the spin-momentum locking,
spin-plasmon manifests itself as coupled charge- and spin-density waves. These waves, accompanying spin-plasmon with
the momentum $\mathbf{q}$, can be characterized by corresponding spatial harmonics of charge- (\ref{rhoDensity}) and
spin-density
\begin{equation}
\mathbf{s}^+_{\mathbf{q}}=\frac12\sum_{\mathbf{p}\gamma\gamma'}\langle
f_{\mathbf{p}+\mathbf{q},\gamma'}|\boldsymbol\sigma|f_{\mathbf{p}\gamma}\rangle
a^+_{\mathbf{p}+\mathbf{q},\gamma'}a_{\mathbf{p}\gamma}\label{s}
\end{equation}
operators.

To relate these quantities to the plasmon operators (\ref{OperatorPlasmon}), we employ the unitary transformation,
connecting operators of electron-hole excitations in noninteracting system
$a^+_{\mathbf{p}+\mathbf{q},\gamma'}a_{\mathbf{p}\gamma}$ with operators of plasmons $Q^+_\mathbf{q}$. For this
purpose, we also need explicit expressions for operators of distorted single-particle excitations in the system with
Coulomb interaction (similarly to the work \cite{Brout}). Denoting creation operator for such single-particle
excitation with total momentum $\mathbf{q}$ and energy $\xi_{\mathbf{p}+\mathbf{q},\gamma'}-\xi_{\mathbf{p}\gamma}$ as
$\eta^+_{\mathbf{pq}\gamma\gamma'}$, we can use the equation-of-motion method in the RPA and, similarly to the case of
plasmons, get the explicit expression for it:
\begin{eqnarray}
\eta^+_{\mathbf{pq}\gamma\gamma'}=\sum_{\mathbf{p}'\tau\tau'}
U^{\gamma\gamma'\tau\tau'}_{\mathbf{p}\mathbf{p}'\mathbf{q}}a^+_{\mathbf{p}'+\mathbf{q},\tau'}a_{\mathbf{p}'\tau},
\label{eta}
\end{eqnarray}
where
\begin{eqnarray}
U^{\gamma\gamma'\tau\tau'}_{\mathbf{p}\mathbf{p}'\mathbf{q}}=\delta_{\mathbf{p}\mathbf{p}'}
\delta_{\gamma\tau}\delta_{\gamma'\tau'}\nonumber\\+\frac{V_q}S \frac{\langle
f_{\mathbf{p}\gamma}|f_{\mathbf{p}+\mathbf{q},\gamma'}\rangle (n_{\mathbf{p}\gamma}-n_{\mathbf{p}+\mathbf{q},\gamma'})}
{1-V_q\Pi(q,\xi_{\mathbf{p}+\mathbf{q},\gamma'}-\xi_{\mathbf{p}\gamma})}\nonumber\\
\times\frac{\langle f_{\mathbf{p}'+\mathbf{q},\tau'}|f_{\mathbf{p}'\tau}\rangle}
{\xi_{\mathbf{p}'\tau}-\xi_{\mathbf{p}'+\mathbf{q},\tau'}-\xi_{\mathbf{p}\gamma}+\xi_{\mathbf{p}+\mathbf{q},\gamma'}+
i\delta}.\label{U}
\end{eqnarray}

The expressions (\ref{OperatorPlasmon}), (\ref{C}), (\ref{eta}) and (\ref{U}) establish the unitary transformation at
given $\mathbf{q}$ from operators of electron-hole excitations in noninteracting system
$a^+_{\mathbf{p}+\mathbf{q},\gamma'}a_{\mathbf{p}\gamma}$ to operators of excitations in Coulomb-interacting system:
plasmons $Q^+_\mathbf{q}$ and single-particle excitations $\eta^+_{\mathbf{pq}\gamma\gamma'}$. We can easily derive the
inverse transformation of the form:
\begin{eqnarray}
a^+_{\mathbf{p}+\mathbf{q},\gamma'}a_{\mathbf{p}\gamma}=(n_{\mathbf{p}\gamma}-n_{\mathbf{p}+\mathbf{q},\gamma'})
\nonumber\\ \times\left\{(C^{\gamma'\gamma}_{\mathbf{p}\mathbf{q}})^*Q^+_\mathbf{q}+\sum_{\mathbf{p}'\tau\tau'}
(U^{\tau\tau'\gamma\gamma'}_{\mathbf{p}'\mathbf{p}\mathbf{q}})^*\eta^+_{\mathbf{p}'\mathbf{q}\tau\tau'}\right\}.
\label{inv}
\end{eqnarray}

According to (\ref{inv}), we can represent the operators (\ref{rhoDensity}) and (\ref{s}) of charge- and spin-density
waves in the form:
\begin{eqnarray}
\rho^+_{\mathbf{q}}=SN_{\mathbf{q}}^*\Pi(q,\Omega_q)Q^+_{\mathbf{q}}+\tilde\rho^+_{\mathbf{q}},\label{rho1}\\
\mathbf{s}^+_{\mathbf{q}}=SN_{\mathbf{q}}^*\boldsymbol\Pi_s(q,\Omega_q)Q^+_{\mathbf{q}}+
\tilde{\mathbf{s}}^+_{\mathbf{q}},\label{s1}
\end{eqnarray}
where the parts $\tilde\rho^+_{\mathbf{q}}$ and $\tilde{\mathbf{s}}^+_{\mathbf{q}}$ are the contributions of
single-particle excitations and are dynamically independent on plasmons. Here, along with the usual susceptibility
(\ref{Pol}), the crossed spin-density susceptibility of the helical liquid \cite{Raghu} has been introduced:
\begin{eqnarray}
\boldsymbol\Pi_s(q,\omega)=\frac1{2S}\sum_{\mathbf{p}\gamma\gamma'}\langle
f_{\mathbf{p}+\mathbf{q},\gamma'}|f_{\mathbf{p}\gamma}\rangle\nonumber\\
\times\langle f_{\mathbf{p}\gamma}|\boldsymbol\sigma|f_{\mathbf{p}+\mathbf{q},\gamma'}\rangle
\frac{n_{\mathbf{p}\gamma}-n_{\mathbf{p}+\mathbf{q},\gamma'}}
{\omega+\xi_{\mathbf{p}\gamma}-\xi_{\mathbf{p}+\mathbf{q},\gamma'}+i\delta}.
\end{eqnarray}

The formulas (\ref{rho1})--(\ref{s1}) show us that the average values of $\rho^+_{\mathbf{q}}$ and
$\mathbf{s}^+_{\mathbf{q}}$ in any state with a definite number of plasmons vanish (similarly to the mean value of
coordinate or momentum in the simplest quantum harmonic oscillator). However, we can calculate their mean squares in
the $n_{\mathbf{q}}$-plasmon state
$|n_{\mathbf{q}}\rangle=[(Q_{\mathbf{q}}^+)^{n_\mathbf{q}}/(n_{\mathbf{q}}!)^{-1/2}]|0\rangle$:
\begin{eqnarray}
\langle\rho_{\mathbf{q}}\rho^+_{\mathbf{q}}\rangle\equiv\langle
n_{\mathbf{q}}|\rho_{\mathbf{q}}\rho^+_{\mathbf{q}}|n_{\mathbf{q}}\rangle-\langle
0|\rho_{\mathbf{q}}\rho^+_{\mathbf{q}}|0\rangle,\label{rho2}\\
\langle s^\perp_{\mathbf{q}}(s^\perp_{\mathbf{q}})^+\rangle\equiv\langle
n_{\mathbf{q}}|s^\perp_{\mathbf{q}}(s^\perp_{\mathbf{q}})^+|n_{\mathbf{q}}\rangle-\langle
0|s^\perp_{\mathbf{q}}(s^\perp_{\mathbf{q}})^+|0\rangle\label{s2}
\end{eqnarray}
(only the in-plane transverse component $s^\perp$ of the spin $\mathbf{s}$ is nonzero upon averaging). Since we are
interested in plasmons only, we have subtracted the vacuum fluctuations of these quantities in the ground state
$|0\rangle$.

\begin{figure}
\includegraphics[width=0.9\columnwidth]{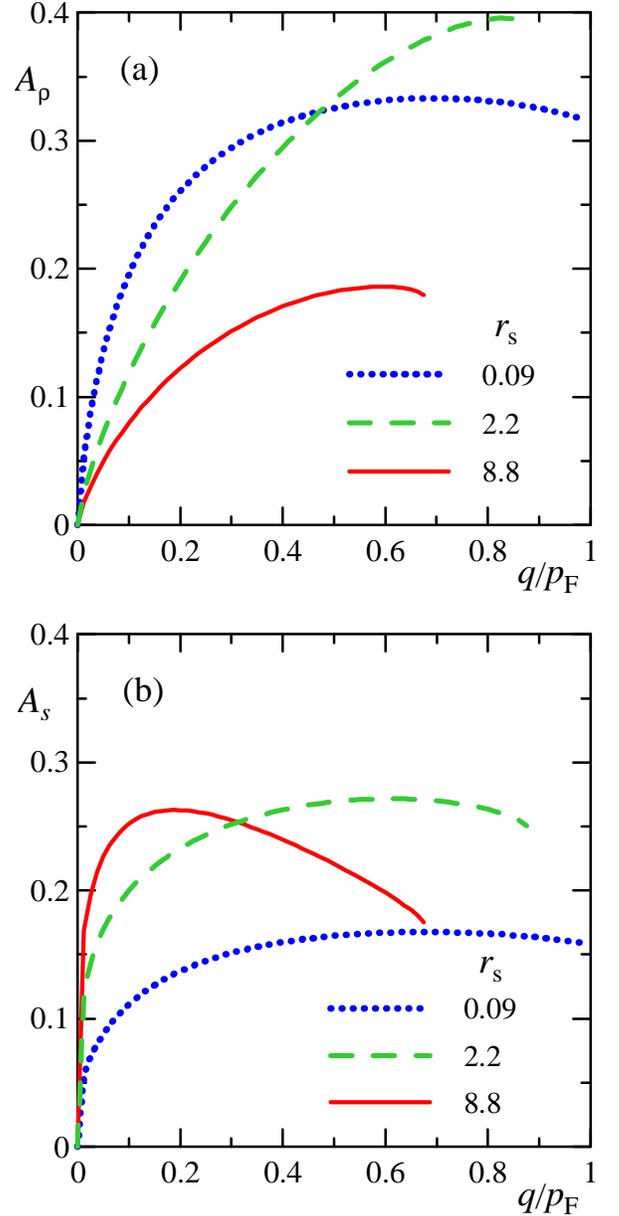}
\caption{Normalized amplitudes of charge-density $A_\rho$ (a) and spin-density $A_s$ (b) waves in many-plasmon state as
function of plasmons momentum $q$ at different values of $r_\mathrm{s}$.}\label{Fig5}
\end{figure}

Using (\ref{rho1})--(\ref{s1}), the mean squares of charge- and spin-density wave amplitudes (\ref{rho2})--(\ref{s2})
can easily be calculated:
\begin{eqnarray}
\langle\rho_{\mathbf{q}}\rho^+_{\mathbf{q}}\rangle=n_{\mathbf{q}}S^2|N_{\mathbf{q}}\Pi(q,\Omega_q)|^2,\\
\langle s^\perp_{\mathbf{q}}(s^\perp_{\mathbf{q}})^+\rangle=n_{\mathbf{q}}S^2|N_{\mathbf{q}}\Pi^\perp_s(q,\Omega_q)|^2.
\end{eqnarray}
We can normalize the amplitudes to obtain dimensionless quantities
$A_\rho(q)=[\langle\rho_{\mathbf{q}}\rho^+_{\mathbf{q}}\rangle/n_{\mathbf{q}}S\rho]^{1/2}$ and $A_s(q)=[\langle
s^\perp_{\mathbf{q}}(s^\perp_{\mathbf{q}})^+\rangle/n_{\mathbf{q}}S\rho]^{1/2}$ ($\rho=p_\mathrm{F}^2/4\pi$ is the
average electron density), plotted in Fig.~\ref{Fig5}. As seen, the amplitudes of charge- and spin-density waves are
close quantitatively at moderate momenta and any~$r_\mathrm{s}$.

In \cite{Raghu}, it was shown that, due to spin-momentum locking, electron density and transverse component of spin
obey an analogue of ``continuity equation''. It requires that
\begin{equation}
\Omega_qA_\rho(q)=2v_\mathrm{F}qA_s(q).
\end{equation}
Our results are in agreement with this equation.

\section{Spin-plasmon scattering}
Nontrivial internal structure of plasmon in Dirac electron gas can reveal itself in a process of its scattering on
external potentials or impurities. For the case of spin-plasmon, it is also interesting to consider its scattering on
magnetic field, acting on electron spins. Since we are interested in spin effects here, we will consider only in-plane
magnetic field, affecting only spins of electrons and not their orbital motion.

Hamiltonian of interaction of electrons in helical liquid with external electric $U(\mathbf{r})$ and magnetic
$\mathbf{H}(\mathbf{r})$ fields are, respectively,
\begin{eqnarray}
H_\mathrm{e}=-e\sum_{\mathbf{p}\mathbf{p}'\gamma\gamma'}U_{\mathbf{p}^\prime-\mathbf{p}}\langle f_{\mathbf{p}^\prime
\gamma^\prime}|f_{\mathbf{p}\gamma}\rangle \, a_{\mathbf{p}^\prime \gamma^\prime}^+a_{\mathbf{p}\gamma},\label{He}\\
H_\mathrm{m}=\mu_\mathrm{B}\sum_{\mathbf{p}\mathbf{p}'\gamma\gamma'}\mathbf{H}_{\mathbf{p}^\prime-\mathbf{p}}\langle
f_{\mathbf{p}^\prime \gamma^\prime}|\boldsymbol\sigma|f_{\mathbf{p}\gamma}\rangle \, a_{\mathbf{p}^\prime
\gamma^\prime}^+a_{\mathbf{p}\gamma},\label{Hm}
\end{eqnarray}
where $U_\mathbf{q}$ and $\mathbf{H}_\mathbf{q}$ are Fourier components of external electric and magnetic fields,
$\mu_\mathrm{B}=e/2mc$ is the Bohr magneton.

To calculate the probability of elastic scattering of the spin-plasmon with initial momentum $\mathbf{q}$ to the state
with the final momentum $\mathbf{q}'$, we can use the Fermi golden rule, corresponding to the Born approximation. Thus,
the differential (with respect to the scattering angle) probabilities of this scattering on electric and magnetic
fields can be presented in the following form:
\begin{eqnarray}
\frac{dw_\mathrm{e}}{d\theta}=\left|V_{\mathbf{q}'-\mathbf{q}}\right|^2
\left|\Phi_\mathrm{e}(q,\theta)\right|^2,\label{Pe1}\\
\frac{dw_\mathrm{e}}{d\theta}=\left|\mathbf{H}_{\mathbf{q}'-\mathbf{q}}\cdot
\mathbf{\Phi}_{\mathrm{m}}(q,\theta)\right|^2,\label{Pm1}
\end{eqnarray}
where $\theta$ is scattering angle (i.e. the angle between $\mathbf{q}$ and $\mathbf{q}'$); $q$ is absolute value of
both $\mathbf{q}$ and $\mathbf{q}'$. Here we have introduced electric $\Phi_\mathrm{e}(q,\theta)$  and  magnetic
$\mathbf{\Phi}_\mathrm{m}(q,\theta)$ form-factors of spin-plasmon.

Using (\ref{He})--(\ref{Hm}), we can calculate the form-factors explicitly:
\begin{eqnarray}
\Phi_\mathrm{e}(q,\theta)=e\sqrt{\frac{q}{\Omega_q^\prime}}\sum_{\mathbf{p}\gamma\gamma^\prime\tau} \left(\langle
f_{\mathbf{p}+\mathbf{q}^\prime,\gamma^\prime}|f_{\mathbf{p}+\mathbf{q},\tau}\rangle\right.\nonumber\\
\left.\times C_{\mathbf{p}\mathbf{q}}^{\tau\gamma}-
C_{\mathbf{p}+\mathbf{q}^\prime-\mathbf{q},\mathbf{q}}^{\gamma^\prime\tau}\langle
f_{\mathbf{p}+\mathbf{q}^\prime-\mathbf{q},\tau}|f_{\mathbf{p}\gamma}\rangle \right)\nonumber\\
\times(C_{\mathbf{p}\mathbf{q}^\prime}^{\gamma^\prime\gamma})^* (n_{\mathbf{p}\gamma}-n_{\mathbf{p}+\mathbf{q}^\prime
\gamma^\prime}),\label{electricFF}\\
\mathbf{\Phi}_\mathrm{m}(q,\theta)=\mu_\mathrm{B}\sqrt{\frac{q}{\Omega_q^\prime}}\sum_{\mathbf{p}\gamma\gamma^\prime\tau}
\left(\langle f_{\mathbf{p}+\mathbf{q}^\prime\gamma^\prime}|\boldsymbol\sigma|f_{\mathbf{p}+\mathbf{q}\tau}\rangle
\right.\nonumber\\ \left.\times
C_{\mathbf{p},\mathbf{q}}^{\tau\gamma}-C_{\mathbf{p}+\mathbf{q}^\prime-\mathbf{q},\mathbf{q}}^{\gamma^\prime\tau}
\langle f_{\mathbf{p}+\mathbf{q}^\prime-\mathbf{q}\tau}|\boldsymbol\sigma|f_{\mathbf{p}\gamma}\rangle \right)\nonumber\\
\times(C_{\mathbf{p}\mathbf{q}^\prime}^{\gamma^\prime\gamma})^* (n_{\mathbf{p}\gamma}-n_{\mathbf{p}+\mathbf{q}^\prime
\gamma^\prime}).
\end{eqnarray}
Here $\Omega_q^\prime=d\Omega_q/dq$ is the derivative of spin-plasmon dispersion law. It is convenient to project the
vector $\mathbf\Phi_\mathrm{m}$ on directions, parallel and perpendicular to the initial plasmon momentum $\mathbf{q}$
to get $\Phi_\mathrm{m}^\parallel$ and $\Phi_\mathrm{m}^\perp$ respectively.

\begin{figure}
\includegraphics[width=\columnwidth]{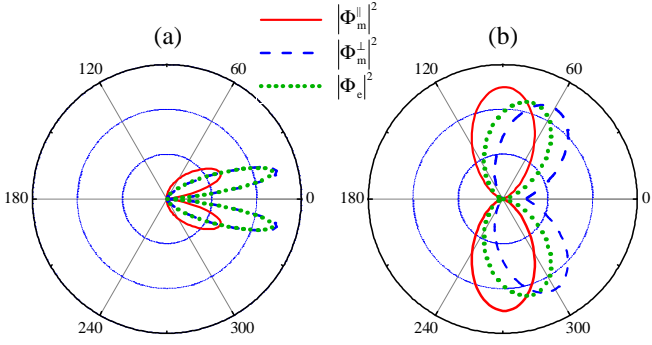}
\caption{Polar graphs of the squared modulus of electric $|\Phi_\mathrm{e}(q,\theta)|^2$ and components
$|\Phi_\mathrm{m}^\parallel(q,\theta)|^2$, $|\Phi_\mathrm{m}^\perp(q,\theta)|^2$ of magnetic form-factor of
spin-plasmon with the momentum $q=0.6p_\mathrm{F}$ at $r_\mathrm{s}=0.09$ (a) and $r_\mathrm{s}=8.8$ (b).}\label{Fig6}
\end{figure}

We consider only the form-factors, revealing the specifics of spin-plasmons, instead of the whole differential
probabilities (\ref{Pe1})--(\ref{Pm1}), also dependent on the form of external field. In Fig.~\ref{Fig6}, angle
dependencies of squared modulus $|\Phi_\mathrm{e}(q,\theta)|^2$, $|\Phi_\mathrm{m}^\parallel(q,\theta)|^2$ and
$|\Phi_\mathrm{m}^\perp(q,\theta)|^2$ of the form-factors at $q=0.6p_\mathrm{F}$ are plotted; these angular
distributions are normalized to unity. In their calculation, we have used the single-band approximation, since undamped
plasmons consist mainly of interband transitions, according to Fig.~\ref{Fig2}. Results for other values of plasmon
momentum are qualitatively the same.

In the case of forward scattering with zero momentum transfer (at $\theta=0$), the external electric field probes the
total charge of plasmon, which is actually zero. Thus, the corresponding electric form-factor $\Phi_\mathrm{e}(q,0)=0$.
Similarly, forward scattering on magnetic field probes the total spin of spin-plasmon, which is directed
perpendicularly to $\mathbf{q}$. Therefore, $\Phi_\mathrm{m}^\parallel(q,0)=0$ and $\Phi_\mathrm{m}^\perp(q,0)\neq0$.
As for backscattering of plasmon, it is strictly prohibited, as for individual massless Dirac electrons \cite{Ando}.

As seen in Fig.~\ref{Fig6}, the form-factors demonstrate two side lobes, rather sharp at small $r_\mathrm{s}$, which
can be considered as a consequence of sharp peaking of the plasmon wave function (see Fig.~\ref{Fig3}). At large
$r_\mathrm{s}$, these lobes are much broader.

In the process of spin-plasmon scattering on some configurations of electric or magnetic fields, we can expect
interplay between spatial structure of these fields and that of the spin-plasmon. Angular distribution of the scattered
plasmons will incorporate both of these factors, according to (\ref{Pe1})--(\ref{Pm1}). Controlling the overlap of
maxima of external potential and spin-plasmon form-factors, one can manipulate spin-plasmon scattering.

Magnetic or nonmagnetic impurities can also create the external field. If the characteristic radius of the impurity
potential is $R$, the transferred momentum will be limited by $R^{-1}$ by the order of magnitude. If $qR\gg1$, the
plasmon scattering on impurities will be suppressed due to proximity to the regime of forward scattering. In the
opposite limit $qR\ll1$, plasmons will be effectively scattered on considerable angles.

\section{Conclusions}
The properties of collective excitations (plasmons or spin-plasmons) in two-dimensional gas of massless Dirac particles
were studied. Two physical realizations of such systems were considered: electron gas in graphene and helical liquid on
the surface of topological insulator. Quantum field-theoretical formalism for comprehensive description of
spin-plasmons as composite Bose particles in the random phase approximation was developed. Internal structure and wave
function of spin-plasmons were studied.

Signatures of spin-momentum locking in helical liquid were considered. In particular, it was shown that excitation of a
spin-plasmon induces the total nonzero spin polarization of the system. Moreover, coupling between charge- and
spin-density waves, accompanying a spin-plasmon, was demonstrated. It was shown that amplitudes of both of these waves
are close by the order of magnitude for spin-plasmons of intermediate momenta. The results of this work can be
confirmed by experiments involving spin-plasmon excitation on the surface of topological insulator and independent
measurements of charge and spin wave amplitudes (one of experiments of this type was proposed in \cite{Raghu}). The
similar effect of coupling between charge and spin appears in electron gas with spin-orbit coupling \cite{Magarill},
but amplitude of spin wave is considerable less in this case than that of charge waves for experimentally relevant
parameters.

Elastic scattering of spin-plasmons in helical liquid on electric and magnetic external fields is considered. Angular
distribution of scattered spin-plasmons depend on both the shape of the potential and the form-factor of the
spin-plasmon, revealing its internal structure. It was shown that, due to the form-factor, the scattering occurs into
two side lobes, while forward and backward scattering is suppressed. One can use this fact to manipulate spin-plasmon
scattering via interplay between plasmon form-factor and form-factor of the external potential. It can be also
concluded that scattering of spin-plasmons on long-range impurities should be very weak.

Coupling between charge and spin waves, demonstrated in this article, can be used for realization of various spintronic
devices. One can perform controllable focusing of spin-plasmon waves and thus create regions with high spin
polarization. Spin-polarized electrons, accumulating in these regions, can diffuse to adjacent electrodes and be used
to drive spin currents (similarly to \cite{Appelbaum}).

The quantum field-theoretical approach, presented in this article, can be used for theoretical description of an
influence of various external factors (impurities, external fields) on plasmons in Dirac electron gas. The problems of
two-dimensional spin-plasmon optics, based on manipulations by inhomogeneities of the system and three-dimensional
environment, can be solved. Also the properties of hybrid modes (plasmon polaritons \cite{Zayats}, plasmon-phonon modes
\cite{LiuWillis,HwangSensarmaDasSarma}, plasmon-hole modes --- plasmarons \cite{Walter}) can be studied using this
approach.

The classical electrodynamic approach based on Maxwell equations and response functions cannot describe quantum
effects, arising when individual plasmons are emitted and adsorbed. Therefore, the quantum field-theoretical approach,
presented in this article, should be especially useful for the problems of quantum plasmonics, which seems to be rather
feasible in graphene-based structures \cite{KoppensChangGarciaAbajo}.

The work was supported by Russian Foundation for Basic Research (grants 11-02-12209-ofi-m and
10-02-92607-KO\underline{\hphantom{a}}a) and by Grant of the President of Russian Federation MK-5288.2011.2. One of the
authors (AAS) acknowledges support from the Dynasty Foundation.

\bibliography{article}

%Merlin.mbs v4.21 2009-07-09.
\begin{thebibliography}{10}%
\makeatletter
\providecommand \@ifxundefined [1]{%
 \ifx #1\undefined \expandafter \@firstoftwo
 \else \expandafter \@secondoftwo
\fi
}%
\providecommand \@ifnum [1]{%
 \ifnum #1\expandafter \@firstoftwo
 \else \expandafter \@secondoftwo
\fi
}%
\providecommand \enquote [1]{``#1''}%
\providecommand \bibnamefont  [1]{#1}%
\providecommand \bibfnamefont [1]{#1}%
\providecommand \citenamefont [1]{#1}%
\providecommand\href[0]{\@sanitize\@href}%
\providecommand\@href[1]{\endgroup\@@startlink{#1}\endgroup\@@href}%
\providecommand\@@href[1]{#1\@@endlink}%
\providecommand \@sanitize [0]{\begingroup\catcode`\&12\catcode`\#12\relax}%
\@ifxundefined \pdfoutput {\@firstoftwo}{%
 \@ifnum{\z@=\pdfoutput}{\@firstoftwo}{\@secondoftwo}%
}{%
 \providecommand\@@startlink[1]{\leavevmode\special{html:<a href="#1">}}%
 \providecommand\@@endlink[0]{\special{html:</a>}}%
}{%
 \providecommand\@@startlink[1]{%
  \leavevmode
  \pdfstartlink
   attr{/Border[0 0 1 ]/H/I/C[0 1 1]}%
   user{/Subtype/Link/A<</Type/Action/S/URI/URI(#1)>>}%
  \relax
 }%
 \providecommand\@@endlink[0]{\pdfendlink}%
}%
\providecommand \url  [0]{\begingroup\@sanitize \@url }%
\providecommand \@url [1]{\endgroup\@href {#1}{\urlprefix}}%
\providecommand \urlprefix [0]{URL }%
\providecommand \Eprint[0]{\href }%
\@ifxundefined \urlstyle {%
  \providecommand \doi [1]{doi:\discretionary{}{}{}#1}%
}{%
  \providecommand \doi [0]{doi:\discretionary{}{}{}\begingroup
  \urlstyle{rm}\Url }%
}%
\providecommand \doibase [0]{http://dx.doi.org/}%
\providecommand \Doi[1]{\href{\doibase#1}}%
\providecommand \bibAnnote [3]{%
  \BibitemShut{#1}%
  \begin{quotation}\noindent
    \textsc{Key:}\ #2\\\textsc{Annotation:}\ #3%
  \end{quotation}%
}%
\providecommand \bibAnnoteFile [2]{%
  \IfFileExists{#2}{\bibAnnote {#1} {#2} {\input{#2}}}{}%
}%
\providecommand \typeout [0]{\immediate \write \m@ne }%
\providecommand \selectlanguage [0]{\@gobble}%
\providecommand \bibinfo [0]{\@secondoftwo}%
\providecommand \bibfield [0]{\@secondoftwo}%
\providecommand \translation [1]{[#1]}%
\providecommand \BibitemOpen[0]{}%
\providecommand \bibitemStop [0]{}%
\providecommand \bibitemNoStop [0]{.\EOS\space}%
\providecommand \EOS [0]{\spacefactor3000\relax}%
\providecommand \BibitemShut [1]{\csname bibitem#1\endcsname}%
%</preamble>
\bibitem{Hasan}%
  \BibitemOpen
  \bibfield{author}{%
  \bibinfo {author} {\bibfnamefont{M~Z}\ \bibnamefont{Hasan}}\ and\ \bibinfo
  {author} {\bibfnamefont{C~L}\ \bibnamefont{Kane}},\ }%
  \bibfield{title}{%
  \enquote{\bibinfo {title} {Topological insulators},}\ }%
  \bibfield{journal}{%
  \bibinfo {journal} {Rev. Mod. Phys}\ }%
  \textbf{\bibinfo {volume} {82}},\ \bibinfo {pages} {3045--3067} (\bibinfo
  {year} {2010})%
  \bibAnnoteFile{NoStop}{Hasan}%
\bibitem{QiZhang}%
  \BibitemOpen
  \bibfield{author}{%
  \bibinfo {author} {\bibfnamefont{X~L}\ \bibnamefont{Qi}}\ and\ \bibinfo
  {author} {\bibfnamefont{S~C}\ \bibnamefont{Zhang}},\ }%
  \bibfield{title}{%
  \enquote{\bibinfo {title} {The quantum spin hall effect and topological
  insulators},}\ }%
  \bibfield{journal}{%
  \bibinfo {journal} {Physics Today}\ }%
  \textbf{\bibinfo {volume} {63}},\ \bibinfo {pages} {33--37} (\bibinfo {year}
  {2010})%
  \bibAnnoteFile{NoStop}{QiZhang}%
\bibitem{QiHughesZhang}%
  \BibitemOpen
  \bibfield{author}{%
  \bibinfo {author} {\bibfnamefont{X~L}\ \bibnamefont{Qi}}, \bibinfo {author}
  {\bibfnamefont{T~L}\ \bibnamefont{Hughes}},\ and\ \bibinfo {author}
  {\bibfnamefont{S~C}\ \bibnamefont{Zhang}},\ }%
  \bibfield{title}{%
  \enquote{\bibinfo {title} {Topological field theory of time-reversal
  invariant insulators},}\ }%
  \bibfield{journal}{%
  \bibinfo {journal} {Phys. Rev. B}\ }%
  \textbf{\bibinfo {volume} {78}},\ \bibinfo {pages} {195424--195467} (\bibinfo
  {year} {2009})%
  \bibAnnoteFile{NoStop}{QiHughesZhang}%
\bibitem{EssinMooreVanderbilt}%
  \BibitemOpen
  \bibfield{author}{%
  \bibinfo {author} {\bibfnamefont{A~M}\ \bibnamefont{Essin}}, \bibinfo
  {author} {\bibfnamefont{J~E}\ \bibnamefont{Moore}},\ and\ \bibinfo {author}
  {\bibfnamefont{D}~\bibnamefont{Vanderbilt}},\ }%
  \bibfield{title}{%
  \enquote{\bibinfo {title} {Magnetoelectric polarizability and axion
  electrodynamics in crystalline insulators},}\ }%
  \bibfield{journal}{%
  \bibinfo {journal} {Phys. Rev. Lett.}\ }%
  \textbf{\bibinfo {volume} {102}},\ \bibinfo {pages} {146805--146809}
  (\bibinfo {year} {2009})%
  \bibAnnoteFile{NoStop}{EssinMooreVanderbilt}%
\bibitem{Chen}%
  \BibitemOpen
  \bibfield{author}{%
  \bibinfo {author} {\bibfnamefont{Y~L}\ \bibnamefont{Chen}}, \bibinfo {author}
  {\bibfnamefont{J~G}\ \bibnamefont{Analytis}}, \bibinfo {author}
  {\bibfnamefont{Z~H}\ \bibnamefont{Chu}}, \bibinfo {author}
  {\bibfnamefont{Z~K}\ \bibnamefont{Liu}}, \bibinfo {author}
  {\bibfnamefont{S~K}\ \bibnamefont{Mo}}, \bibinfo {author}
  {\bibfnamefont{X~L}\ \bibnamefont{Qi}}, \bibinfo {author}
  {\bibfnamefont{H~J}\ \bibnamefont{Zhang}}, \bibinfo {author}
  {\bibfnamefont{D~H}\ \bibnamefont{Lu}}, \bibinfo {author}
  {\bibfnamefont{X}~\bibnamefont{Dai}}, \bibinfo {author}
  {\bibfnamefont{Z}~\bibnamefont{Fang}}, \bibinfo {author} {\bibfnamefont{S~C}\
  \bibnamefont{Zhang}}, \bibinfo {author} {\bibfnamefont{I~R}\
  \bibnamefont{Fisher}}, \bibinfo {author}
  {\bibfnamefont{Z}~\bibnamefont{Hussain}},\ and\ \bibinfo {author}
  {\bibfnamefont{Z~X}\ \bibnamefont{Shen}},\ }%
  \bibfield{title}{%
  \enquote{\bibinfo {title} {Large gap topological insulator bi2te3 with a
  single dirac cone on the surface},}\ }%
  \bibfield{journal}{%
  \bibinfo {journal} {Science}\ }%
  \textbf{\bibinfo {volume} {325}},\ \bibinfo {pages} {178--182} (\bibinfo
  {year} {2009})%
  \bibAnnoteFile{NoStop}{Chen}%
\bibitem{Hsieh}%
  \BibitemOpen
  \bibfield{author}{%
  \bibinfo {author} {\bibfnamefont{D}~\bibnamefont{Hsieh}}, \bibinfo {author}
  {\bibfnamefont{Y}~\bibnamefont{Xia}}, \bibinfo {author}
  {\bibfnamefont{D}~\bibnamefont{Qian}}, \bibinfo {author}
  {\bibfnamefont{L}~\bibnamefont{Wray}}, \bibinfo {author}
  {\bibfnamefont{F}~\bibnamefont{Dil}, \bibfnamefont{Z~Hand~Meier}}, \bibinfo
  {author} {\bibfnamefont{J}~\bibnamefont{Osterwalder}}, \bibinfo {author}
  {\bibfnamefont{J~G}\ \bibnamefont{Patthey}, \bibfnamefont{L~Checkelsky}},
  \bibinfo {author} {\bibfnamefont{NP}~\bibnamefont{Ong}}, \bibinfo {author}
  {\bibfnamefont{H}~\bibnamefont{Fedorov}, \bibfnamefont{AV~Lin}}, \bibinfo
  {author} {\bibfnamefont{A}~\bibnamefont{Bansil}}, \bibinfo {author}
  {\bibfnamefont{D}~\bibnamefont{Grauer}}, \bibinfo {author}
  {\bibfnamefont{Y~S}\ \bibnamefont{Hor}}, \bibinfo {author}
  {\bibfnamefont{R~J}\ \bibnamefont{Cava}},\ and\ \bibinfo {author}
  {\bibfnamefont{M~Z}\ \bibnamefont{Hasan}},\ }%
  \bibfield{title}{%
  \enquote{\bibinfo {title} {First observation of spin-momentum helical locking
  in bi2se3 and bi2te3, demonstration of topological-order at 300k and a
  realization of topological-transport-regime},}\ }%
  \bibfield{journal}{%
  \bibinfo {journal} {Nature}\ }%
  \textbf{\bibinfo {volume} {460}},\ \bibinfo {pages} {1101--1105} (\bibinfo
  {year} {2009})%
  \bibAnnoteFile{NoStop}{Hsieh}%
\bibitem{Xia}%
  \BibitemOpen
  \bibfield{author}{%
  \bibinfo {author} {\bibfnamefont{Y}~\bibnamefont{Xia}}, \bibinfo {author}
  {\bibfnamefont{L}~\bibnamefont{Wray}}, \bibinfo {author}
  {\bibfnamefont{D}~\bibnamefont{Qian}}, \bibinfo {author}
  {\bibfnamefont{D}~\bibnamefont{Hsieh}}, \bibinfo {author}
  {\bibfnamefont{A}~\bibnamefont{Pal}}, \bibinfo {author}
  {\bibfnamefont{H}~\bibnamefont{Lin}}, \bibinfo {author}
  {\bibfnamefont{A}~\bibnamefont{Bansil}}, \bibinfo {author}
  {\bibfnamefont{D}~\bibnamefont{Grauer}}, \bibinfo {author}
  {\bibfnamefont{Y}~\bibnamefont{Hor}}, \bibinfo {author}
  {\bibfnamefont{R}~\bibnamefont{Cava}},\ and\ \bibinfo {author}
  {\bibfnamefont{M~Z}\ \bibnamefont{Hasan}},\ }%
  \bibfield{title}{%
  \enquote{\bibinfo {title} {Discovery (theoretical prediction and experimental
  observation) of a large-gap topological-insulator class with spin-polarized
  single-dirac-cone on the surface},}\ }%
  \bibfield{journal}{%
  \bibinfo {journal} {Nature Phys.}\ }%
  \textbf{\bibinfo {volume} {5}},\ \bibinfo {pages} {398--402} (\bibinfo {year}
  {2009})%
  \bibAnnoteFile{NoStop}{Xia}%
\bibitem{Burkov}%
  \BibitemOpen
  \bibfield{author}{%
  \bibinfo {author} {\bibfnamefont{A~A}\ \bibnamefont{Burkov}}\ and\ \bibinfo
  {author} {\bibfnamefont{D~G}\ \bibnamefont{Hawthorn}},\ }%
  \bibfield{title}{%
  \enquote{\bibinfo {title} {Spin and charge transport on the surface of a
  topological insulator},}\ }%
  \bibfield{journal}{%
  \bibinfo {journal} {Phys. Rev. Lett.}\ }%
  \textbf{\bibinfo {volume} {105}},\ \bibinfo {pages} {066802--066806}
  (\bibinfo {year} {2010})%
  \bibAnnoteFile{NoStop}{Burkov}%
\bibitem{Cucler}%
  \BibitemOpen
  \bibfield{author}{%
  \bibinfo {author} {\bibfnamefont{D}~\bibnamefont{Culcer}}, \bibinfo {author}
  {\bibfnamefont{E~H}\ \bibnamefont{Hwang}}, \bibinfo {author}
  {\bibfnamefont{D~T}\ \bibnamefont{Stanescu}},\ and\ \bibinfo {author}
  {\bibfnamefont{S}~\bibnamefont{Das~Sarma}},\ }%
  \bibfield{title}{%
  \enquote{\bibinfo {title} {Two-dimensional surface charge transport in
  topological insulators},}\ }%
  \bibfield{journal}{%
  \bibinfo {journal} {Rev. Mod. Phys.}\ }%
  \textbf{\bibinfo {volume} {82}},\ \bibinfo {pages} {155457--155474} (\bibinfo
  {year} {2010})%
  \bibAnnoteFile{NoStop}{Cucler}%
\bibitem{YokoyamaTanakaNagaosa}%
  \BibitemOpen
  \bibfield{author}{%
  \bibinfo {author} {\bibfnamefont{T}~\bibnamefont{Yokoyama}}, \bibinfo
  {author} {\bibfnamefont{Y}~\bibnamefont{Tanaka}}, ,\ and\ \bibinfo {author}
  {\bibfnamefont{N}~\bibnamefont{Nagaosa}},\ }%
  \bibfield{title}{%
  \enquote{\bibinfo {title} {Giant spin rotation in the junction between a
  normal metal and a quantum spin hall system},}\ }%
  \bibfield{journal}{%
  \bibinfo {journal} {Phys. Rev. Lett.}\ }%
  \textbf{\bibinfo {volume} {102}},\ \bibinfo {pages} {166801--166805}
  (\bibinfo {year} {2009})%
  \bibAnnoteFile{NoStop}{YokoyamaTanakaNagaosa}%
\bibitem{Raghu}%
  \BibitemOpen
  \bibfield{author}{%
  \bibinfo {author} {\bibfnamefont{S}~\bibnamefont{Raghu}}, \bibinfo {author}
  {\bibfnamefont{S~B}\ \bibnamefont{Chung}}, \bibinfo {author}
  {\bibfnamefont{X~L}\ \bibnamefont{Qi}},\ and\ \bibinfo {author}
  {\bibfnamefont{S~C}\ \bibnamefont{Zhang}},\ }%
  \bibfield{title}{%
  \enquote{\bibinfo {title} {Collective modes of a helical liquid},}\ }%
  \bibfield{journal}{%
  \bibinfo {journal} {Phys. Rev. Lett.}\ }%
  \textbf{\bibinfo {volume} {104}},\ \bibinfo {pages} {116401--116405}
  (\bibinfo {year} {2010})%
  \bibAnnoteFile{NoStop}{Raghu}%
\bibitem{Appelbaum}%
  \BibitemOpen
  \bibfield{author}{%
  \bibinfo {author} {\bibfnamefont{I}~\bibnamefont{Appelbaum}}, \bibinfo
  {author} {\bibfnamefont{H~D}\ \bibnamefont{Drew}},\ and\ \bibinfo {author}
  {\bibfnamefont{M~S}\ \bibnamefont{Fuhrer}},\ }%
  \bibfield{title}{%
  \enquote{\bibinfo {title} {Proposal for a topological plasmon spin
  rectifier},}\ }%
  \bibfield{journal}{%
  \bibinfo {journal} {Appl. Phys. Lett.}\ }%
  \textbf{\bibinfo {volume} {98}},\ \bibinfo {pages} {023103--023107} (\bibinfo
  {year} {2011})%
  \bibAnnoteFile{NoStop}{Appelbaum}%
\bibitem{Karch}%
  \BibitemOpen
  \bibfield{author}{%
  \bibinfo {author} {\bibfnamefont{A}~\bibnamefont{Karch}},\ }%
  \enquote{\bibinfo {title} {Surface plasmons and topological insulators},}\
  (\bibinfo {year} {2011}),\
  \Eprint{http://arxiv.org/abs/1104.4125v2}{arXiv:1104.4125v2}%
  \bibAnnoteFile{NoStop}{Karch}%
\bibitem{Graphene1}%
  \BibitemOpen
  \bibfield{author}{%
  \bibinfo {author} {\bibfnamefont{K~S}\ \bibnamefont{Novoselov}}, \bibinfo
  {author} {\bibfnamefont{A~K}\ \bibnamefont{Geim}}, \bibinfo {author}
  {\bibfnamefont{S~V}\ \bibnamefont{Morozov}}, \bibinfo {author}
  {\bibfnamefont{D}~\bibnamefont{Jiang}}, \bibinfo {author}
  {\bibfnamefont{Y}~\bibnamefont{Zhang}}, \bibinfo {author}
  {\bibfnamefont{S~V}\ \bibnamefont{Dubonos}}, \bibinfo {author}
  {\bibfnamefont{I~V}\ \bibnamefont{Grigorieva}},\ and\ \bibinfo {author}
  {\bibfnamefont{A~A}\ \bibnamefont{Firsov}},\ }%
  \bibfield{title}{%
  \enquote{\bibinfo {title} {Electric field effects in atomically thin carbon
  films},}\ }%
  \bibfield{journal}{%
  \bibinfo {journal} {Science}\ }%
  \textbf{\bibinfo {volume} {306}},\ \bibinfo {pages} {666--670} (\bibinfo
  {year} {2004})%
  \bibAnnoteFile{NoStop}{Graphene1}%
\bibitem{Graphene2}%
  \BibitemOpen
  \bibfield{author}{%
  \bibinfo {author} {\bibfnamefont{K~S}\ \bibnamefont{Novoselov}}, \bibinfo
  {author} {\bibfnamefont{A~K}\ \bibnamefont{Geim}}, \bibinfo {author}
  {\bibfnamefont{S~V}\ \bibnamefont{Morozov}}, \bibinfo {author}
  {\bibfnamefont{D}~\bibnamefont{Jiang}}, \bibinfo {author}
  {\bibfnamefont{M~I}\ \bibnamefont{Katsnelson}}, \bibinfo {author}
  {\bibfnamefont{I~V}\ \bibnamefont{Grigorieva}}, \bibinfo {author}
  {\bibfnamefont{S~V}\ \bibnamefont{Dubonos}},\ and\ \bibinfo {author}
  {\bibfnamefont{A~A}\ \bibnamefont{Firsov}},\ }%
  \bibfield{title}{%
  \enquote{\bibinfo {title} {Two-dimensional gas of massless dirac fermions in
  graphene},}\ }%
  \bibfield{journal}{%
  \bibinfo {journal} {Nature}\ }%
  \textbf{\bibinfo {volume} {438}},\ \bibinfo {pages} {197--201} (\bibinfo
  {year} {2005})%
  \bibAnnoteFile{NoStop}{Graphene2}%
\bibitem{CastroNeto}%
  \BibitemOpen
  \bibfield{author}{%
  \bibinfo {author} {\bibfnamefont{A~H}\ \bibnamefont{Castro~Neto}}, \bibinfo
  {author} {\bibfnamefont{F}~\bibnamefont{Guinea}}, \bibinfo {author}
  {\bibfnamefont{N~M~R}\ \bibnamefont{Peres}}, \bibinfo {author}
  {\bibfnamefont{K~S}\ \bibnamefont{Novoselov}},\ and\ \bibinfo {author}
  {\bibfnamefont{A~K}\ \bibnamefont{Geim}},\ }%
  \bibfield{title}{%
  \enquote{\bibinfo {title} {Electronic properties of graphene},}\ }%
  \bibfield{journal}{%
  \bibinfo {journal} {Rev. Mod. Phys.}\ }%
  \textbf{\bibinfo {volume} {81}},\ \bibinfo {pages} {109--162} (\bibinfo
  {year} {2009})%
  \bibAnnoteFile{NoStop}{CastroNeto}%
\bibitem{KotovUchoaPereiraCastroNetoGuinea}%
  \BibitemOpen
  \bibfield{author}{%
  \bibinfo {author} {\bibfnamefont{V~N}\ \bibnamefont{Kotov}}, \bibinfo
  {author} {\bibfnamefont{B}~\bibnamefont{Uchoa}}, \bibinfo {author}
  {\bibfnamefont{V~M}\ \bibnamefont{Pereira}}, \bibinfo {author}
  {\bibfnamefont{A~H}\ \bibnamefont{Castro~Neto}},\ and\ \bibinfo {author}
  {\bibfnamefont{F}~\bibnamefont{Guinea}},\ }%
  \enquote{\bibinfo {title} {Electron-electron interactions in graphene:
  Current status and perspectives},}\  (\bibinfo {year} {2011}),\
  \Eprint{http://arxiv.org/abs/1012.3484v1}{arXiv:1012.3484v1}%
  \bibAnnoteFile{NoStop}{KotovUchoaPereiraCastroNetoGuinea}%
\bibitem{HwangDasSarma}%
  \BibitemOpen
  \bibfield{author}{%
  \bibinfo {author} {\bibfnamefont{E~H}\ \bibnamefont{Hwang}}\ and\ \bibinfo
  {author} {\bibfnamefont{S}~\bibnamefont{Das~Sarma}},\ }%
  \bibfield{title}{%
  \enquote{\bibinfo {title} {Dielectric function, screening, and plasmons in 2d
  graphene},}\ }%
  \bibfield{journal}{%
  \bibinfo {journal} {Phys. Rev. B}\ }%
  \textbf{\bibinfo {volume} {75}},\ \bibinfo {pages} {205418--672} (\bibinfo
  {year} {2007})%
  \bibAnnoteFile{NoStop}{HwangDasSarma}%
\bibitem{WunchStauberSolsGuinea}%
  \BibitemOpen
  \bibfield{author}{%
  \bibinfo {author} {\bibfnamefont{B}~\bibnamefont{Wunsch}}, \bibinfo {author}
  {\bibfnamefont{T}~\bibnamefont{Stauber}}, \bibinfo {author}
  {\bibfnamefont{F}~\bibnamefont{Sols}},\ and\ \bibinfo {author}
  {\bibfnamefont{F}~\bibnamefont{Guinea}},\ }%
  \bibfield{title}{%
  \enquote{\bibinfo {title} {Dynamical polarization of graphene at finite
  doping},}\ }%
  \bibfield{journal}{%
  \bibinfo {journal} {New J Phys}\ }%
  \textbf{\bibinfo {volume} {8}},\ \bibinfo {pages} {318--333} (\bibinfo {year}
  {2006})%
  \bibAnnoteFile{NoStop}{WunchStauberSolsGuinea}%
\bibitem{BludovVasilevskiyPeres}%
  \BibitemOpen
  \bibfield{author}{%
  \bibinfo {author} {\bibfnamefont{Y~u~V}\ \bibnamefont{Bludov}}, \bibinfo
  {author} {\bibfnamefont{M~I}\ \bibnamefont{Vasilevskiy}},\ and\ \bibinfo
  {author} {\bibfnamefont{N~M~R}\ \bibnamefont{Peres}},\ }%
  \bibfield{title}{%
  \enquote{\bibinfo {title} {Mechanism for graphene-based optoelectronic
  switches by tuning surface plasmon-polaritons in monolayer},}\ }%
  \bibfield{journal}{%
  \bibinfo {journal} {Europhysics Letters}\ }%
  \textbf{\bibinfo {volume} {92}},\ \bibinfo {pages} {68001--68006} (\bibinfo
  {year} {2010})%
  \bibAnnoteFile{NoStop}{BludovVasilevskiyPeres}%
\bibitem{HwangSensarmaDasSarma}%
  \BibitemOpen
  \bibfield{author}{%
  \bibinfo {author} {\bibfnamefont{E~H}\ \bibnamefont{Hwang}}, \bibinfo
  {author} {\bibfnamefont{R}~\bibnamefont{Sensarma}},\ and\ \bibinfo {author}
  {\bibfnamefont{S}~\bibnamefont{Das~Sarma}},\ }%
  \bibfield{title}{%
  \enquote{\bibinfo {title} {Plasmon-phonon coupling in graphene},}\ }%
  \bibfield{journal}{%
  \bibinfo {journal} {Phys. Rev. B}\ }%
  \textbf{\bibinfo {volume} {82}},\ \bibinfo {pages} {195406--195411} (\bibinfo
  {year} {2010})%
  \bibAnnoteFile{NoStop}{HwangSensarmaDasSarma}%
\bibitem{LiuWillis}%
  \BibitemOpen
  \bibfield{author}{%
  \bibinfo {author} {\bibfnamefont{Y~u}\ \bibnamefont{Liu}}\ and\ \bibinfo
  {author} {\bibfnamefont{R~F}\ \bibnamefont{Willis}},\ }%
  \bibfield{title}{%
  \enquote{\bibinfo {title} {Plasmon-phonon strongly-coupled mode in epitaxial
  graphene},}\ }%
  \bibfield{journal}{%
  \bibinfo {journal} {Phys Rev B.}\ }%
  \textbf{\bibinfo {volume} {81}},\ \bibinfo {pages} {081406--081410} (\bibinfo
  {year} {2010})%
  \bibAnnoteFile{NoStop}{LiuWillis}%
\bibitem{KoppensChangGarciaAbajo}%
  \BibitemOpen
  \bibfield{author}{%
  \bibinfo {author} {\bibfnamefont{F~H~L}\ \bibnamefont{Koppens}}, \bibinfo
  {author} {\bibfnamefont{D~E}\ \bibnamefont{Chang}},\ and\ \bibinfo {author}
  {\bibfnamefont{F~J~G}\ \bibnamefont{de~Abajo}},\ }%
  \enquote{\bibinfo {title} {Graphene plasmonics: A platform for strong
  light-matter interaction},}\  (\bibinfo {year} {2011}),\
  \Eprint{http://arxiv.org/abs/1104.2068}{arXiv:1104.2068}%
  \bibAnnoteFile{NoStop}{KoppensChangGarciaAbajo}%
\bibitem{Zhang}%
  \BibitemOpen
  \bibfield{author}{%
  \bibinfo {author} {\bibfnamefont{H}~\bibnamefont{Zhang}}, \bibinfo {author}
  {\bibfnamefont{C~X}\ \bibnamefont{Liu}}, \bibinfo {author}
  {\bibfnamefont{X~L}\ \bibnamefont{Qi}}, \bibinfo {author}
  {\bibfnamefont{X}~\bibnamefont{Dai}}, \bibinfo {author}
  {\bibfnamefont{Z}~\bibnamefont{Fang}},\ and\ \bibinfo {author}
  {\bibfnamefont{S~C}\ \bibnamefont{Zhang}},\ }%
  \bibfield{title}{%
  \enquote{\bibinfo {title} {Topological insulators in bi2se3, bi2te3 and
  sb2te3 with a single dirac cone on the surface},}\ }%
  \bibfield{journal}{%
  \bibinfo {journal} {Nature Phys.}\ }%
  \textbf{\bibinfo {volume} {5}},\ \bibinfo {pages} {438--442} (\bibinfo {year}
  {2009})%
  \bibAnnoteFile{NoStop}{Zhang}%
\bibitem{Sawada}%
  \BibitemOpen
  \bibfield{author}{%
  \bibinfo {author} {\bibfnamefont{K}~\bibnamefont{Sawada}}, \bibinfo {author}
  {\bibfnamefont{K~A}\ \bibnamefont{Brueckner}}, \bibinfo {author}
  {\bibfnamefont{N}~\bibnamefont{Fukuda}},\ and\ \bibinfo {author}
  {\bibfnamefont{R}~\bibnamefont{Brout}},\ }%
  \bibfield{title}{%
  \enquote{\bibinfo {title} {Correlation enery of an electron gas at heigh
  density: plasma oscillations},}\ }%
  \bibfield{journal}{%
  \bibinfo {journal} {Phys. Rev.}\ }%
  \textbf{\bibinfo {volume} {108}},\ \bibinfo {pages} {507--514} (\bibinfo
  {year} {1957})%
  \bibAnnoteFile{NoStop}{Sawada}%
\bibitem{Apenko}%
  \BibitemOpen
  \bibfield{author}{%
  \bibinfo {author} {\bibfnamefont{S~M}\ \bibnamefont{Apenko}}, \bibinfo
  {author} {\bibfnamefont{D~A}\ \bibnamefont{Kirzhnits}},\ and\ \bibinfo
  {author} {\bibfnamefont{Yu~E}\ \bibnamefont{Lozovik}},\ }%
  \bibfield{title}{%
  \enquote{\bibinfo {title} {On the validity of the 1/n-expansion},}\ }%
  \bibfield{journal}{%
  \bibinfo {journal} {Phys. Lett. A}\ }%
  \textbf{\bibinfo {volume} {92}},\ \bibinfo {pages} {107--109} (\bibinfo
  {year} {1982})%
  \bibAnnoteFile{NoStop}{Apenko}%
\bibitem{Brout}%
  \BibitemOpen
  \bibfield{author}{%
  \bibinfo {author} {\bibfnamefont{R}~\bibnamefont{Brout}},\ }%
  \bibfield{title}{%
  \enquote{\bibinfo {title} {Correlation energy of a high-density gas: Plasma
  coordinates},}\ }%
  \bibfield{journal}{%
  \bibinfo {journal} {Phys. Rev.}\ }%
  \textbf{\bibinfo {volume} {108}},\ \bibinfo {pages} {515--517} (\bibinfo
  {year} {1957})%
  \bibAnnoteFile{NoStop}{Brout}%
\bibitem{Ando}%
  \BibitemOpen
  \bibfield{author}{%
  \bibinfo {author} {\bibfnamefont{T}~\bibnamefont{Ando}}, \bibinfo {author}
  {\bibfnamefont{T}~\bibnamefont{Nakanishi}},\ and\ \bibinfo {author}
  {\bibfnamefont{R}~\bibnamefont{Saito}},\ }%
  \bibfield{title}{%
  \enquote{\bibinfo {title} {Berry's phase and absence of back scattering in
  carbon nanotubes},}\ }%
  \bibfield{journal}{%
  \bibinfo {journal} {J. Phys. Soc. Jpn.}\ }%
  \textbf{\bibinfo {volume} {67}},\ \bibinfo {pages} {2857--2862} (\bibinfo
  {year} {1998})%
  \bibAnnoteFile{NoStop}{Ando}%
\bibitem{Magarill}%
  \BibitemOpen
  \bibfield{author}{%
  \bibinfo {author} {\bibfnamefont{L~I}\ \bibnamefont{Magarill}}, \bibinfo
  {author} {\bibfnamefont{A~V}\ \bibnamefont{Chaplik}},\ and\ \bibinfo {author}
  {\bibfnamefont{M~V}\ \bibnamefont{\'{E}ntin}},\ }%
  \bibfield{title}{%
  \enquote{\bibinfo {title} {Spin-plasmon oscillations of the two-dimensional
  electron gas},}\ }%
  \bibfield{journal}{%
  \bibinfo {journal} {JETP}\ }%
  \textbf{\bibinfo {volume} {92}},\ \bibinfo {pages} {153--158} (\bibinfo
  {year} {2001})%
  \bibAnnoteFile{NoStop}{Magarill}%
\bibitem{Zayats}%
  \BibitemOpen
  \bibfield{author}{%
  \bibinfo {author} {\bibfnamefont{A~V}\ \bibnamefont{Zayats}}, \bibinfo
  {author} {\bibfnamefont{I~I}\ \bibnamefont{Smolyaninov}},\ and\ \bibinfo
  {author} {\bibfnamefont{A~A}\ \bibnamefont{Maradudin}},\ }%
  \bibfield{title}{%
  \enquote{\bibinfo {title} {Nano-optics of surface plasmon polaritons},}\ }%
  \bibfield{journal}{%
  \bibinfo {journal} {Phys. Rep.}\ }%
  \textbf{\bibinfo {volume} {408}},\ \bibinfo {pages} {131--314} (\bibinfo
  {year} {2005})%
  \bibAnnoteFile{NoStop}{Zayats}%
\bibitem{Walter}%
  \BibitemOpen
  \bibfield{author}{%
  \bibinfo {author} {\bibfnamefont{A~L}\ \bibnamefont{Walter}}, \bibinfo
  {author} {\bibfnamefont{A}~\bibnamefont{Bostwick}}, \bibinfo {author}
  {\bibfnamefont{K-J}\ \bibnamefont{Jeon}}, \bibinfo {author}
  {\bibfnamefont{F}~\bibnamefont{Speck}}, \bibinfo {author}
  {\bibfnamefont{M}~\bibnamefont{Ostler}}, \bibinfo {author}
  {\bibfnamefont{T}~\bibnamefont{Seyller}}, \bibinfo {author}
  {\bibfnamefont{L}~\bibnamefont{Moreschini}}, \bibinfo {author}
  {\bibfnamefont{Y~J}\ \bibnamefont{Chang}}, \bibinfo {author}
  {\bibfnamefont{M}~\bibnamefont{Polini}}, \bibinfo {author}
  {\bibfnamefont{R}~\bibnamefont{Asgari}}, \bibinfo {author}
  {\bibfnamefont{A~H}\ \bibnamefont{MacDonald}}, \bibinfo {author}
  {\bibfnamefont{E}~\bibnamefont{Horn}},\ and\ \bibinfo {author}
  {\bibfnamefont{E}~\bibnamefont{Rotenberg}},\ }%
  \bibfield{title}{%
  \enquote{\bibinfo {title} {Effective screening and the plasmaron bands in
  graphene},}\ }%
  \bibfield{journal}{%
  \bibinfo {journal} {Phys. Rev. B 84}\ }%
  \textbf{\bibinfo {volume} {84}},\ \bibinfo {pages} {085410} (\bibinfo {year}
  {2011})%
  \bibAnnoteFile{NoStop}{Walter}%
\end{thebibliography}%

\end{document}